\documentclass{article}

\usepackage[final]{neurips_2024}

\usepackage[utf8]{inputenc} 
\usepackage[T1]{fontenc}    
\usepackage{hyperref}       
\usepackage{url}            
\usepackage{booktabs}       
\usepackage{amsfonts}       
\usepackage{nicefrac}       
\usepackage{microtype}      
\usepackage{xcolor}         
\usepackage{graphicx} 
\usepackage{amsmath}
\usepackage{amssymb}
\usepackage{amssymb}
\usepackage{wrapfig}
\usepackage{authblk} 

\makeatletter
\renewcommand\AB@affilsepx{, \protect\Affilfont}
\makeatother

\title{Neural decoding from stereotactic EEG: accounting for electrode variability across subjects}

\author[1, *]{Georgios Mentzelopoulos}
\author[1]{Evangelos Chatzipantazis}
\author[2]{Ashwin G. Ramayya}
\author[2]{Michelle J. Hedlund}
\author[2]{Vivek P. Buch}
\author[1, 3]{Kostas Daniilidis}
\author[1]{Konrad P. Kording}
\author[1, *]{Flavia Vitale}
\affil[1]{University of Pennsylvania}
\affil[2]{Stanford University}
\affil[3]{Archimedes, Athena RC}

\begin{document}
\renewcommand{\thefootnote}{\textasteriskcentered}
\footnotetext[1]{Contact: gment@upenn.edu, vitalef@pennmedicine.upenn.edu \\ \hphantom{~~~~~~} Project page: \href{https://gmentz.github.io/seegnificant}{https://gmentz.github.io/seegnificant}}

\maketitle

\begin{abstract}
Deep learning based neural decoding from stereotactic electroencephalography (sEEG) would likely benefit from scaling up both dataset and model size.
To achieve this, combining data across multiple subjects is crucial. However, in sEEG cohorts, each subject has a variable number of electrodes placed at distinct locations in their brain, solely based on clinical needs. Such heterogeneity in electrode number/placement poses a significant challenge for data integration, since there is no clear correspondence of the neural activity recorded at distinct sites between individuals. Here we introduce seegnificant: a training framework and architecture that can be used to decode behavior across subjects using sEEG data. We tokenize the neural activity within electrodes using convolutions and extract long-term temporal dependencies between tokens using self-attention in the time dimension. 
{The 3D location of each electrode is then mixed with the tokens, followed by another self-attention in the electrode dimension to extract effective spatiotemporal neural representations}. Subject-specific heads are then used for downstream decoding tasks. Using this approach, we construct a multi-subject model trained on the combined data from 21 subjects performing a behavioral task. We demonstrate that our model is able to decode the trial-wise response time of the subjects during the behavioral task solely from neural data. We also show that the neural representations learned by pretraining our model across individuals can be transferred in a few-shot manner to new subjects. This work introduces a scalable approach towards sEEG data integration for multi-subject model training, paving the way for cross-subject generalization for sEEG decoding.  

\end{abstract}

\section{Introduction} \label{sec:introduction}

Deep learning has revolutionized many fields ranging from natural language processing \citep{Vaswani2017}, to computer vision \citep{Dosovitskiy2020}, and neural decoding \citep{Azabou2023}. Advances in these fields have shown that performance and generalization benefit from scaling up both datasets and model size, underscoring the importance of integrating data across multiple individuals. Recent work by \cite{Azabou2023} showed that such approaches can achieve excellent generalization in the motor decoding domain, by combining invasive microelectrode array recordings from multiple non-human primates. Despite their promise, however, such approaches have not been used to build scalable decoders for human brain recordings collected with stereotactic electroenchephalography (sEEG). Compared to other approaches (e.g., ECoG, microelectrodes), sEEG is minimally invasive and is currently the gold-standard clinical tool for invasive electrophysiology in humans. Therefore, building strong neural decoding models using sEEG promises to be medically relevant. 

There are several challenges associated with decoding from sEEG data. A major barrier is the heterogeneity of the sEEG cohorts. SEEG subjects are most often patients suffering from pharmacoresistant epilepsy that are implanted with sEEG electrodes as part of their medical care. Each subject gets implanted with a variable number of electrodes in specific locations in their brain, solely based on clinical needs \citep{Kovac2017}. This heterogeneity makes it hard to identify any correspondence between the neural activity recorded across different subjects. 
This lack of correspondence poses a significant challenge to combining data across subjects in a meaningful way, resulting in most research groups resorting to within-subject models \citep{Wu2024, Angrick2021, Petrosyan2022, Meng2021}. While somewhat effective, single-subject approaches neither scale nor generalize, suggesting a need to effectively integrate multi-subject data.  

Here we introduce seegnificant: a scalable framework that can be used to decode behavior from sEEG recordings using multi-session, multi-subject data. Our approach uses a unified feature extraction backbone, designed to extract global behaviorally relevant neural representations that are shared across subjects. It uses personalized task-heads, tailored to the idiosyncrasies of individual participants, that can be used for downstream decoding tasks. To account for the heterogeneous electrode placement across individuals, we use a convolutional tokenizer that operates within electrodes individually. The tokenized latents are then processed by self-attention within the time dimension to capture long-term temporal dependencies within electrodes. Using a novel positional encoding scheme, we then imprint the latents of each electrode with information about their 3D location in the brain, using their MNI coordinates, a standarized 3D coordinate system used to localize brain regions \citep{talairach1988, collins1994}. The spatially aware latents are then processed by another self-attention that operates within the electrode dimension to capture long-range dependencies across electrodes. The resulting latents, are then compressed into neural representations that capture the spatiotemporal dependencies within and between electrodes that are used for behavioral decoding.

To evaluate our approach, we combine data from 21 subjects, across 29 recording sessions, that performed a behavioral reaction time task, while their sEEG was simultaneously recorded, totaling more that 3600 behavioral trials and 100 electrode-hours (number of electrodes $\times$ number of recording hours) of sEEG recording. We demonstrate that using our approach, we are able to extract global neural representations, capable of decoding the response time of the participants during the behavioral task from neural data. Using a leave-one-out cross validation scheme, we also show that by pretraining our model on large amounts of data, we can transfer the learned neural representations to new subjects with minimal training examples. Our work introduces a novel framework designed to train models on multi-session, multi-subject sEEG data for behavioral decoding applications in a scalable way. 

Our contributions can be summarized as:

\begin{itemize}
    \item \textit{A framework for multi-subject training based on sEEG.} We present a novel approach to training transformer based models for neural decoding based on the combined data of multi-subject, multi-session sEEG datasets.
    \item \textit{Pre-trained models for behavioral decoding based on sEEG.} We trained a multi-session, multi-subject model for response time decoding based on sEEG that can be finetuned to new subjects. We will make the model and code publicly available as a resource for the community.
\end{itemize}

\section{Related Work} \label{sec:related_work}

\subsection{Neural decoding using sEEG}
Over the past decade, significant progress has been made in the field of neural decoding using sEEG. \cite{Angrick2021} demonstrated effective real-time speech synthesis based on sEEG using a linear discriminant analysis approach, and \cite{Petrosyan2022} showed effective speech decoding using a convolutional architecture. \cite{Meng2021} identified discriminative features for decoding speech perception and overt and imagined speech production from sEEG using a non-deep learning model. \cite{Wu2022} showed that grasp force can be successfully decoded from sEEG signals using a CNN+RNN decoder and subsequently performed a comparative study where they explored the efficacy of using a variety of convolutional architectures to classify different movement based on the sEEG neural activity \citep{Wu2023, Wu2024}. 
Unfortunately, all aforementioned works conduct experiments on datasets containing only a few subjects (12 or less, except \cite{Wu2023}) and build within-subject models, with no clear vision towards generalization across subjects. Diverging from within-subject approaches, in this work we combine data from many subjects to build models capable of extracting global neural representations that generalize {across subjects}. Contrary to previous works which used CNNs and/or RNNs, we build our models using the transformer architecture, which has been shown to excel in capturing {long-range} dependencies when trained on large and diverse datasets.

\subsection{Transformer architectures for neural data}

Transformers have attracted a lot of attention in the analysis of non-invasive {neural signals}. In the context of non-invasive electroencephalography (EEG), multiple works use the attention mechanism for various decoding tasks such as motor-imagery, visual decoding, and emotional recognition \citep{Tang2024, Kan2023, Li2023, Liu2022, Lan2020}. \cite{Song2023} achieved SOTA performance in a visual decoding task using a convolutional tokenizer that operates within both time and electrode dimensions and is then processed using self-attention. Using the same tokenization approach, \cite{NeuroGPT} built neuro-GPT, a foundational model for non-invasive EEG based on transformers. However, non-invasive EEG uses a predefined number of electrodes placed at consistent locations of the scalp across subjects. This approach provides a well-defined correspondence across the neural activity recorded between subjects. In sEEG, the number and placement of electrodes across subjects is highly variable and inconsistent between subjects, complicating efforts to translate approaches from EEG to sEEG data.

In the context of invasive microelectrode recordings, prior works have attempted to use transformers for neural decoding with success. \cite{Ye2021} introduced the Neural Data Transformers (NDT) to model neural population activity as an alternative to RNN models \citep{Glaser2020, Sussillo2016}.
\cite{Trung2022} extended the NDT model to process data in both the electrode and time dimensions. However, all aforementioned works trained single-subjects models. \cite{Azabou2023} diverged from single subject approaches by combining data from multiple non-human primates and achieved excellent decoding performance in a variety of different motor decoding tasks. However microelectrode recordings have a well-defined tokenization scheme based on extracellular neuronal spikes, which are not detectable by mm-scale sEEG electrodes. Thus, it is unclear how spike-based tokenization approaches can be meaningfully translated to sEEG.

\subsection{Shared trunk architectures}

{Shared trunk architectures are a popular choice for multi task learning applications. They follow a simple outline: a global feature extractor, whose parameters are shared across tasks, followed by task-specific branches, with parameters specific to each task \citep{Crawshaw2020}. They have been successfully used in multiple domains including facial recognition \citep{Zhang2014}, instance-aware semantic segmentation \citep{Dai2015}, image retrieval \citep{Zhao2018} and classification \citep{Liu2019CV}. Their effectiveness stems from their ability to extract shared representations that are common to all tasks. Combining data from multiple different tasks also helps mitigate the large scale data requirements for effective training of deep networks. In this work, while we are not dealing with multiple different tasks, we use a shared trunk architecture to model inter-person differences between subjects in our cohort. Our network is composed of a backbone, common to all subjects, that extracts global representations and subject-specific heads that tailor the model's output to the unique statistical profile of each subject.}

\section{{Methodology}} \label{sec:approach}

SEEG is a neural recording modality capable of recording local field potentials (LFPs) that reflect the coordinated activation of hundreds of thousands of neurons in the vicinity of an electrode, providing a mesoscale measurement of brain activity \citep{Saez2023}. Due to the distributed nature of sEEG, subjects typically get implanted with tens to hundreds of electrodes in widespread locations across their brain. The resulting data are multi-variate time series that are typically epoched around a behavioral stimulus of interest (e.g. a visual stimulus presented during each trial of a behavioral task). 

Neural activity recorded via sEEG across the electrodes of an individual are, of course, not independent. The captured recordings represent a conversation between thousands of neurons in distributed brain networks. The challenge is to find a way to interpret those recordings in the context of this broader conversation, instead of interpreting them in isolation \citep{Azabou2023}. This challenge is exacerbated when trying to integrate neural recordings across different subjects, each of which has a variable number of electrodes distributed across different brain regions. This is equivalent to monitoring the neural conversation with several microphones placed in different brain locations unique to each subject. For each monitored neural population, we do not know the identity or the functional tuning (what stimuli they respond to).

\begin{figure}[h]
    \centering
    \includegraphics[scale=0.50]{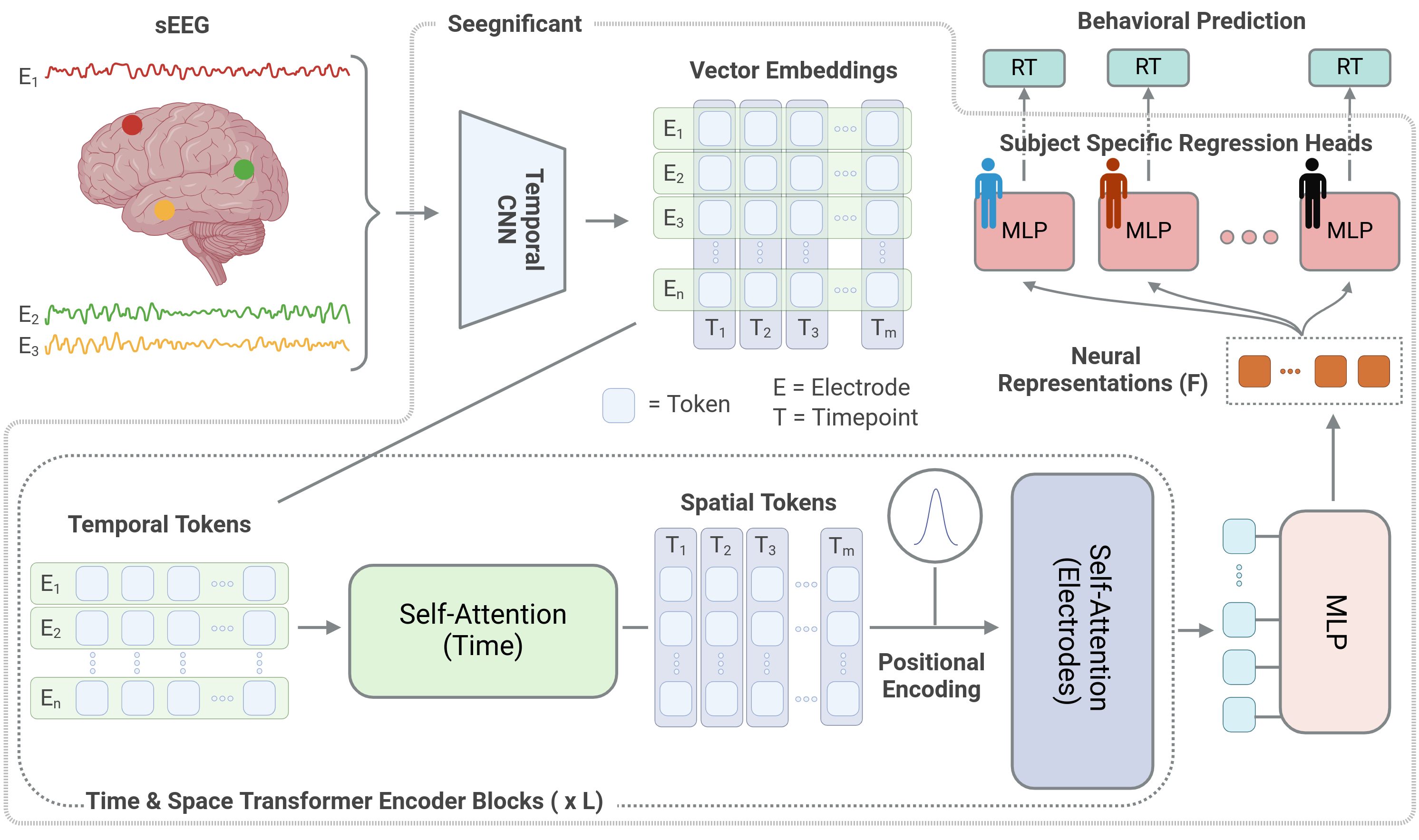} 
    \caption{\textit{Outline of our network architecture}. The sEEG signals are converted to vector embeddings through temporal convolutions and then processed by sequential self-attention operations in the time and electrode dimensions in an alternating fashion. The latents are compressed and projected through subject-specific task heads to obtain behavioral predictions. }
    \label{networkArchitectrure}
\end{figure}

\subsection{Signal processing} \label{sec:signal_processing}

Motivated by the lack of correspondence between the neural activity recorded between subjects, we sought to identify the subset of electrodes across subjects that respond to a behavioral stimulus of interest (i.e., share the same functional tuning in the neural conversation) while subjects perform a behavioral task. This can be achieved using the methodology introduced by \cite{Coon2016} and refined by \cite{Paraskevopoulou2021}. This methodology identifies electrodes whose high-$\gamma$ band activity (narrowband neural activity with frequency content between 70-150 Hz) is modulated by the behavioral stimulus of interest. 
We note that high-$\gamma$ band activity is used because it has been associated with the instantaneous firing of neuronal populations in the vicinity of an electrode \citep{Miller2009} and, thus, it likely represents neural computation that is relevant to the behavioral stimulus of interest.

As described by \cite{Paraskevopoulou2021}, separately for each electrode, the neural signals are filtered to the high-$\gamma$ band and the envelope is extracted using the Hilbert transform. The high-$\gamma$ envelopes are then epoched around the behavioral stimulus of interest and separated into a baseline period (short time window prior to stimulus) and task period (short time window after the stimulus). The median high-$\gamma$ envelope of the task ($Median_{task}$) and baseline ($Median_{baseline}$) periods are computed across the trials of the behavioral task. Then for each electrode, the signal-to-noise ratio (SNR) is calculated as 
\begin{equation}
    SNR = \frac{\sigma^{2}(Median_{task})}{\sigma^{2}(Median_{baseline})},
\end{equation}
where $\sigma^{2}(\cdot)$ refers to the variance of $(\cdot)$. Electrodes whose high-$\gamma$ band activity is significantly modulated by the behavioral stimulus can then be identified using a bootstrap randomization test and subsequently be used for decoding applications. 

The implementation details of this analysis can be found in Appendix \ref{data_collection_details}. 

\subsection{Network architecture}

\subsubsection{Tokenization} 

To combat the inhomogeneity of the number/placement of electrodes across different individuals, we designed a convolutional tokenizer that operates on the voltage traces of each electrode separately. Instead of performing a two-dimensional convolution across both the time and electrode dimension, which would blend information between the dissimilar electrodes that are randomly placed across individuals, we perform a one-dimensional convolution across the temporal dimension only. This way, each electrode gets assigned a $K$-dimensional learnable embedding for chunks of time, defined by the length of the convolutional kernel. 

Specifically, let $x_{e} \in R^{T_{trial}}$ denote the LFP recorded by electrode $e$ for a given trial of length $T_{trial}$. Let $k_{i}$ denote the $i^{th}$ convolutional kernel, from a collection of $K$ kernels. Our approach returns learnable vector embedding $z_{e} \in R^{T_{trial} \times K}$ by performing convolutions of the form,
\begin{equation}
    z_{e, i} = x_{e} * k_{i}, \text{ for } i \in 0, 1, ... ,  K,
\end{equation}
on $x_{e}$ ({convolutional kernel weights are shared across all the electrodes of all subjects}). After temporal convolutions, batch normalization is applied to the $K$ features extracted by the convolution followed by average pooling along the time dimension (which reduces the time dimension from $T_{trial}$ to $T$), returning $z_{e} \in R^{T \times K}$.

Stacking the results of these operations across electrodes, we obtain the latent $z \in R^{E \times T \times K}$, where $E$ denotes the number of electrodes, $T$ is the number of time samples, and $K$ is the number of features extracted by the convolutions. By representing our data in this way, our network is capable of handling inputs with varying numbers of electrodes and time lengths. We do note, however, that the time resolution (i.e., sampling rate) needs to be fixed for the convolutions to be meaningful across subjects. 

\subsubsection{Capturing long-range spatiotemporal dependencies} \label{section:attn}

The latent $z \in R^{E \times T \times K}$ semantically represents the cumulative neural activity of all electrodes of a subject, for a given trial. It can also be conceptualized as a separable dual tokenization across the time and electrode dimensions, with electrode-latents $z_{e} \in R^{T \times K}$ and time-latents $z_{t} \in R^{E \times K}$. This separability, enables us to process the vector embeddings separately in the time and electrode dimension, giving us a significant computational advantage to subsequent self-attention layers, whose computational complexity scales quadratically with sequence length. 

\textbf{Attention in the time dimension.} We first interrogate our data for long-term temporal dependencies within electrodes. To do so, for each electrode separately, we arrange the latents $z_{e} \in R^{T \times K}$ into a sequence of $T$ tokens, where each token has dimensionality $K$. The tokens are then projected into equally shaped queries: $Q = W_{q}z_{e}$, keys: $K = W_{k}z_{e}$, and values: $V = W_{v}z_{e}$, and processed using self-attention,
\begin{equation} \label{attn}
    \text{Attention}(Q, K, V) = \text{softmax}(\frac{QK^{T}}{{\sqrt{d_{k}}}}) V.
\end{equation}
For this operation we used the standard transformer block \citep{Vaswani2017}, preceding the attention operation with normalization layers and following with a feed-forward network. This operation is parallelized across electrodes, and we stack the results to obtain latents of the form $z_{int} \in R^{E \times T \times K}$. 

\textbf{Spatial positional encoding.} 
To account for the inconsistency of electrode locations across individuals, there is a need to inject information about the spatial location of each electrode  into our model. To do so, we used radial basis functions to capture the information about each electrode location in the brain based on its MNI coordinates \citep{MacDonald2000, talairach1988}. 
We discretize the space of each coordinate, into $n$ bins, with midpoints $\mu_{0}, \mu_{1}, \dots, \mu_{n-1} $. For each midpoint, we centered $m$ univariate gaussians of the form, 

\begin{equation}
    f_{i, j}(s) = \frac{1}{\sqrt{2\pi\sigma_{j}^2}} \exp\left(-\frac{(s - \mu_{i})^2}{2\sigma_{j}^2}\right),
\end{equation}

with variances $\sigma_{j}^{2}$, for $i = 0, 1, \dots, n-1$ and  $j = 0, 1, \dots, m-1$. Separately for the $x$, $y$, and $z$ coordinates, we compute $f_{i, j}(x)$, $f_{i, j}(y)$, and $f_{i, j}(z)$ $\forall$ $i, j$, which we then concatenate into a vector $p \in R^{3 \cdot m \cdot n}$, which represents the positional encoding of an electrode. The positional encodings for each electrode are then projected to $K$ dimensions (using a linear layer) and added to the latents $z_{int} \in R^{E \times T \times K}$. 

\textbf{Attention in the space dimension.} Having captured temporal dependencies within electrodes in our data, and having "stamped" the latents with their spatial positional encodings, we then use self-attention again, this time in the electrode dimension to capture {long-range} dependencies between the neural activity across electrodes in our data. The latents $z_{int} \in R^{E \times T \times K}$ are arranged, separately for each timepoint, into $z_{t} \in R^{E \times K}$ as a sequence of $E$ tokens of dimension $K$, which are projected into equally shaped queries: $Q = W_{q}z_{t}$, keys: $K = W_{k}z_{t}$, and values: $V = W_{v}z_{t}$. The tokens are processed with self-attention using equation \ref{attn}. Again, the standard transformer block is used, with appropriate pre-normalization followed by a feed-forward network. This time, the operation is parallelized across timepoints, and the results are stacked to obtain latents of the form $z_{l} \in R^{E \times T \times K}$.

Multiple layers of self-attention in time and self-attention in space can be stacked back-to-back to enhance the fitting ability of the network, if necessary. Leveraging the versatility of transformers to accept inputs of varying lengths, the attention operation in the both time and space, as described above, does not constrain our model to accept inputs of fixed length in terms of the number of electrodes, or timepoints. In the case where inputs do have unequal number of electrodes or timepoints, computations can be efficiently parallelized by masking tokens that correspond to padded electrodes and timepoints during the self-attention operations.

\textbf{Extracting global features}. In behavioral neuroscience, experimenters are usually interested in decoding few behavioral variables, such as a response time or hit rate to a given stimulus \citep{Posner1990, ANT}. With this in mind, the information contained in the latent $z_{l} \in R^{E \times T \times K}$, will ultimately need to be compressed into a single value. Therefore, following the self-attention operations, we unroll the latents $z_{l}$ and, using a feed-forward network, we project them to a low dimensional neural representation $F \in R^{d}$, where $d$ denotes the final number of the extracted features, with  {$d \ll E \cdot T \cdot K$}, that will be used for the downstream decoding tasks.  

\subsubsection{Personalizing to individual subjects}

Our architecture is built around the idea of extracting neural representations that are common across sEEG subjects. We also, however, need to take into account that performance to behavioral tasks is intrinsically different between individuals \citep{Fozard1994, Davranche2006, Green2003, Der2006}. Therefore, we designed a separate task head for each individual, composed of a shallow feed-forward network, that maps the extracted neural representations $F$ to the behavioral outcome of a given trial.

\begin{figure}[h]
    \centering
    \includegraphics[scale=0.40]{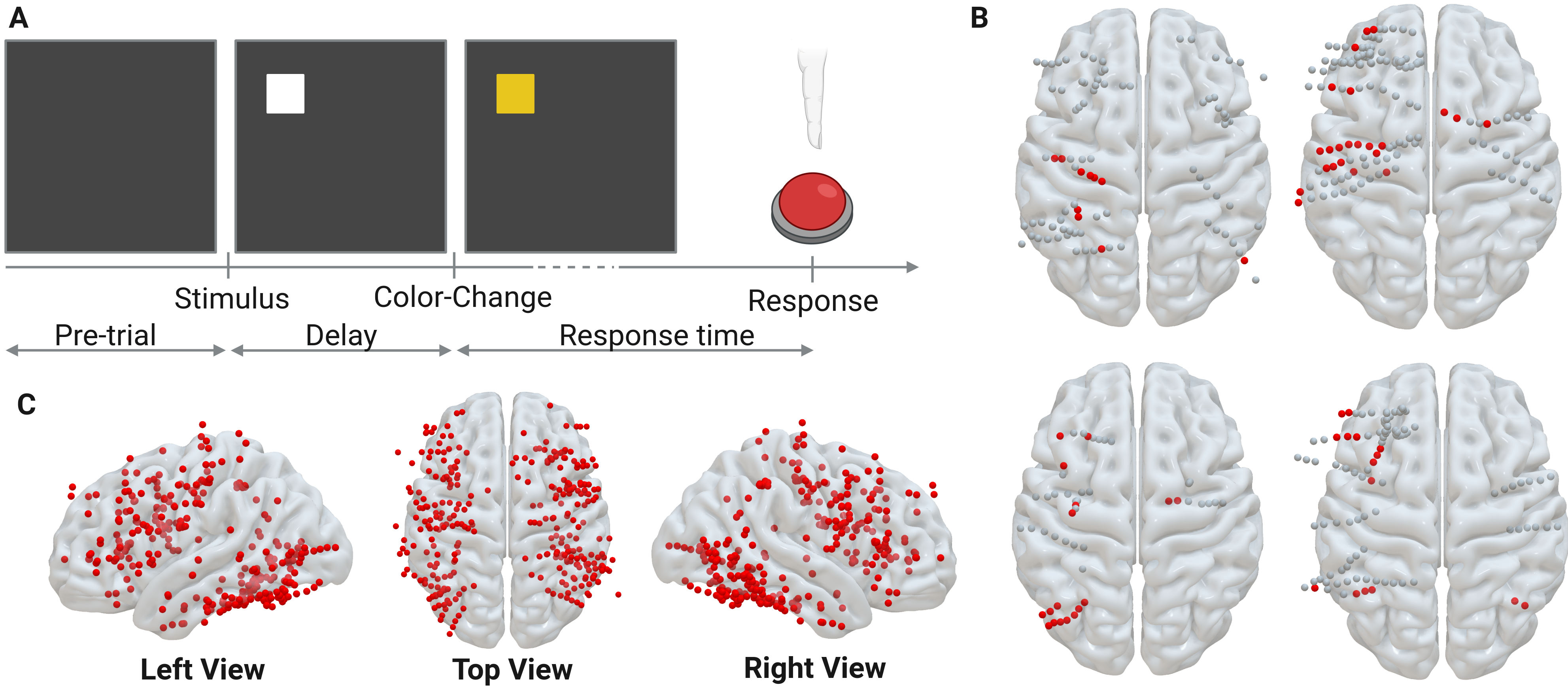} 
    \caption{\textit{Overview of behavioral experiment.} \textbf{A}. Schematic of the color change detection task. \textbf{B}. {Electrode placement projected onto the MNI brain template for four example subjects in our cohort. Red dots show electrodes used for model training; grey dots show electrodes excluded from model training (see section \ref{sec:signal_processing}). \textbf{C}. Electrodes used for model training across all subjects in our cohort projected on an MNI brain template.}}
    \label{experimental_setup}
\end{figure}

\section{Experiments} \label{sec:experiments}

In this section, we test and validate our approach for multi-session, across-subject behavioral decoding using sEEG on a large and diverse cohort of sEEG subjects performing a behavioral task.

\subsection{Experimental setup} \label{sec:dataset_and_experimental_setup}

 {\textbf{Dataset.}} The core design constraint of our approach is to enable the unified training of our architecture (see Fig. \ref{networkArchitectrure}) on a diverse, multi-session, multi-subject sEEG dataset despite the inhomogeneity of electrode number/placement in each subject. Towards that goal, we built a large and diverse dataset of subjects that performed a repetitive reaction time task, while their sEEG was simultaneously recorded. 
Contrary to most studies, which collected a few hundred behavioral trials from few subjects, \textit{we collected more than 3600 behavioral trials and more than 100 electrode-hours of recording from 21 subjects over 29 recording sessions}. {Our dataset is composed of 13 females and 8 males, with ages ranging from 16 to 57 years old, with unique electrode number/placement for each subject, solely based on clinical needs. Across subjects electrodes span white and grey matter; cortical, subcortical, and deep structures (see Fig. \ref{experimental_setup}).} 

 {\textbf{Behavioral task.}} Subjects performed the color-change detection task (see Fig. \ref{experimental_setup}A \& Appendix \ref{data_collection_details}). In each trial, a visual stimulus was presented. After a variable foreperiod delay, the stimulus changed color. At that time, the participant responded by pressing a button as fast as they could. \textit{Our goal was to decode the trial-wise response time of the subjects using their sEEG}. 

\textbf{Design choices}. Throughout all experiments, we use the neural data from a window of [0, 1500] msec after the stimulus color change to train models and make behavioral predictions. All models were trained with $N_{t}=105$ temporal tokens within each electrode and $N_{e} \in [3, 28]$ electrode tokens within each timepoint, which varied based on the number of electrodes for each participant. 
The dimensionality of each token was fixed to $K=2$. All training details are provided in Appendix \ref{ModelTrainingDetails}.

\subsection{Training within-subject models} \label{sec:ss_training}

To test whether our modeling approach would be able to decode the trial-wise response time from sEEG data, we began with a within-subject approach and tested whether our architecture was capable of decoding the response times for each subject. We trained a separate model for each subject in our cohort, combining data across sessions for participants that had multiple behavioral sessions. The experimental setup and the preprocessing of the sEEG data was kept the same for all subject. \textit{Across the 21 single-subject models, the average test set $R^2$ was 0.30 $\pm$ 0.05 (mean $\pm$ sem)}.

\subsection{Training a multi-session, multi-subject model} \label{ms_training}

To investigate whether training on more data, despite the heterogeneity (see Fig. \ref{experimental_setup}B, C), would improve response time decoding performance, we trained a unified model on the combined neural data of all the behavioral trials across all participants in our dataset. This effectively increased the number of training samples 21-fold compared to the data available for the single-subject models. 

\begin{wrapfigure}{r}{0.65\textwidth} 
    \centering
    \includegraphics[scale=0.55]{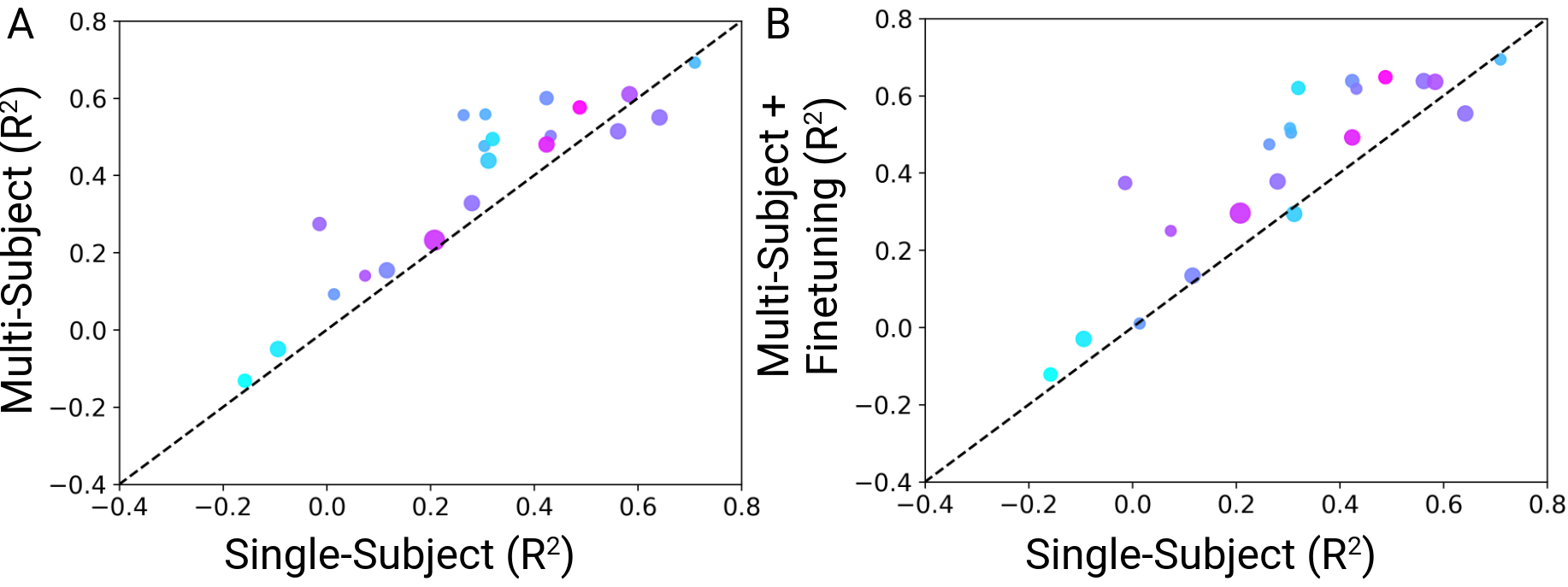} 
    \caption{\textit{Comparing decoding performance between the single-subject and multi-subject models for each subject.} \textbf{A}. Single-subject vs multi-subject model. \textbf{B}. Single-subject vs finetuned, multi-subject model. Circle size denotes the number of trials  and color the number of electrodes (cyan to  magenta represents ascending order)}
    \label{fig:ss_vs_ms}
\end{wrapfigure}

This model achieved a test set $R^2$ of 0.54 $\pm$ 0.01 (mean $\pm$ sem) on the combined data of all participants. 
 {The root mean square error (RMSE) for all predicted response times in the test
set, across subjects, was RMSE = 82 $\pm$ 2 (mean $\pm$ sem) msec. For reference, the mean response
time in the test set, across subjects, is 410 $\pm$ 5 (mean $\pm$ sem) msec, which is an order of magnitude greater than the RMSE.}
Since the response time profiles of each subject are distinct, we also evaluated the model performance within-subjects. \textit{Across subjects, the average per-subject test set $R^2$ was 0.39 $\pm$ 0.05 (mean $\pm$ sem)}. 
Compared to training on single-subjects, the multi-subject training approach boosted decoding performance for our proposed architecture by $\Delta R^{2} = 0.09$, on average.  
We then compared the performance of the single-subject models to that of the multi-subject model, head to head (see Fig. \ref{fig:ss_vs_ms}A). For the majority of subjects, there was a clear performance boost from multi-subject training. 

We then explored whether finetuning the multi-session, multi-subject model to individual subjects would further boost the performance gains, compare to the single-subject models. 
\textit{The finetuned models for each subject achieved an average test set $R^2$ of 0.41 $\pm$ 0.05, across subjects}, indicating that finetuning the multi-session, multi-subject model to single subjects can further boost the performance gains. Importantly, finetuning the unified model boosted the performance gains of the multi-subject model compared to the single-subject models by  {$\Delta R^{2} = 0.11$}, on average (see Fig. \ref{fig:ss_vs_ms}B for per-subject head to head comparisons). These results suggest that sEEG-based neural decoding benefits from multi-subject joint model training, despite the heterogeneity of electrode number/placement across subjects, {demonstrating the power of multi-subject approaches compared to single-subject ones.}

\subsection{Transferring to new subjects} \label{transfering_to_new_subjects}

Having shown that training our model across diverse, multi-subject data boosts decoding performance, we were interested in identifying whether the representations learned by pretraining our models on multiple subjects can efficiently be transferred to new subjects, unseen from the model during training. This is important in any real-world clinical scenario, where only a few number of behavioral trials can be collected from a subject. To test this, we employed a leave-one-out cross validation approach. We trained 21 models, each of which was trained on the combined data of all subjects but one. The average test set $R^2$ of the multi-subject models was 0.48 $\pm$ 0.006 (mean $\pm$ sem) across all trials of all subject. This indicates that our training framework is robust to data variations and that our model did not depend on the data of any specific subject to train effectively. 

\begin{wrapfigure}{r}{0.65\textwidth} 
    \centering
    \includegraphics[scale=0.55]{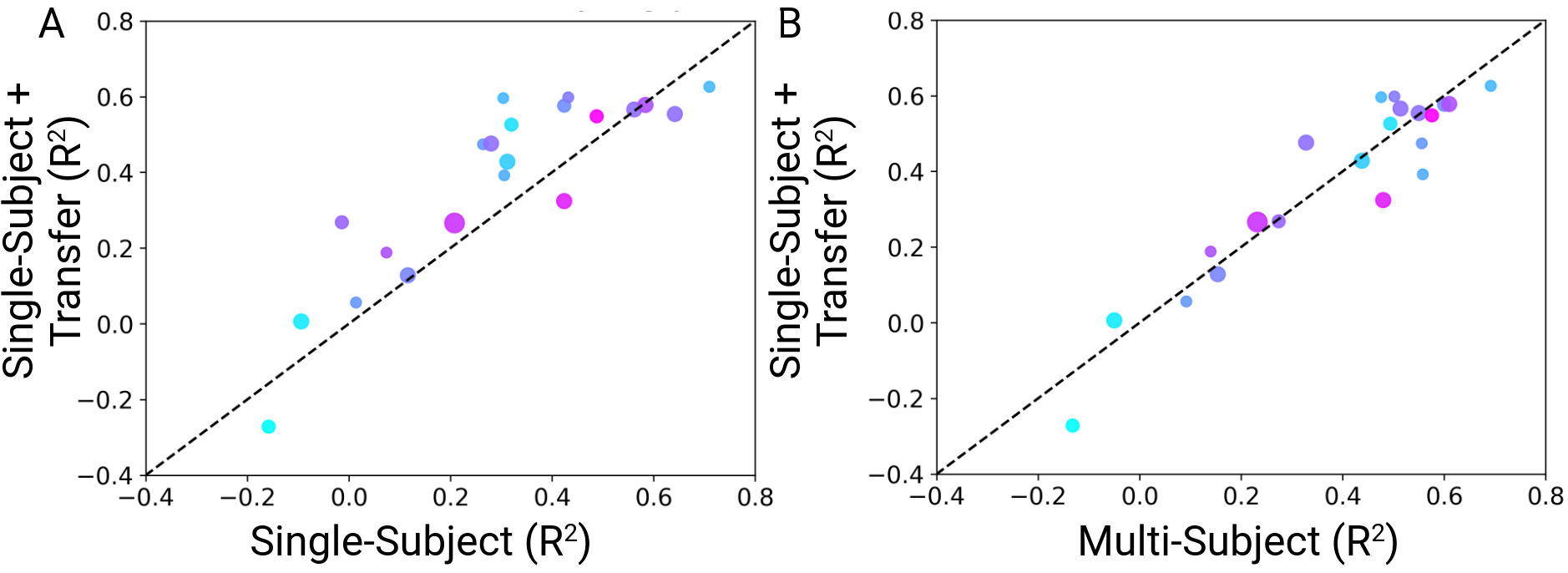} 
    \caption{\textit{Comparing decoding performance between} \textbf{(A)} Transferred single-subject finetuned models vs  single-subject models trained from scratch, and \textbf{(B)} Transferred single-subject finetuned models vs the multi-session, multi-subject model. }
    \label{fig:transfering_to_new_subjects}
\end{wrapfigure}

The weights of the pretrained models were then used as the basis for finetuning on the data of each left-out participant, unseen from the models during training. \textit{Across subjects, the  models trained on other subjects and transferred to a new one achieved an average test set $R^{2}$ of 0.38 $\pm$ 0.05 (mean $\pm$ sem)}. This is a clear improvement to training single-subject models from scratch and comparable to training multi-session, multi-subject models. Specifically, the transferred single-subject models showed a per-subject decoding performance increase $\Delta R^{2} = 0.08$ compared to our single-subject models trained from scratch (see Fig. \ref{fig:transfering_to_new_subjects}A) and a per-subject decoding performance decrease $\Delta R^{2} = -0.01$, compared to the multi-session, multi-subject model (Fig. \ref{fig:transfering_to_new_subjects}B)).

This suggests that pretraining models on diverse, multi-subject data and using their weights as a finetune to transfer the learned representations to new subjects is almost as effective as training on the combined data of all subjects. Given the significant computational benefit of the transferring approach, its implementation in real world medical scenarios might be more likely.

\subsection{Comparison with baselines} \label{sec:baseline_comparisons}

\begin{figure}[h]
    \centering
    \includegraphics[scale=0.35]{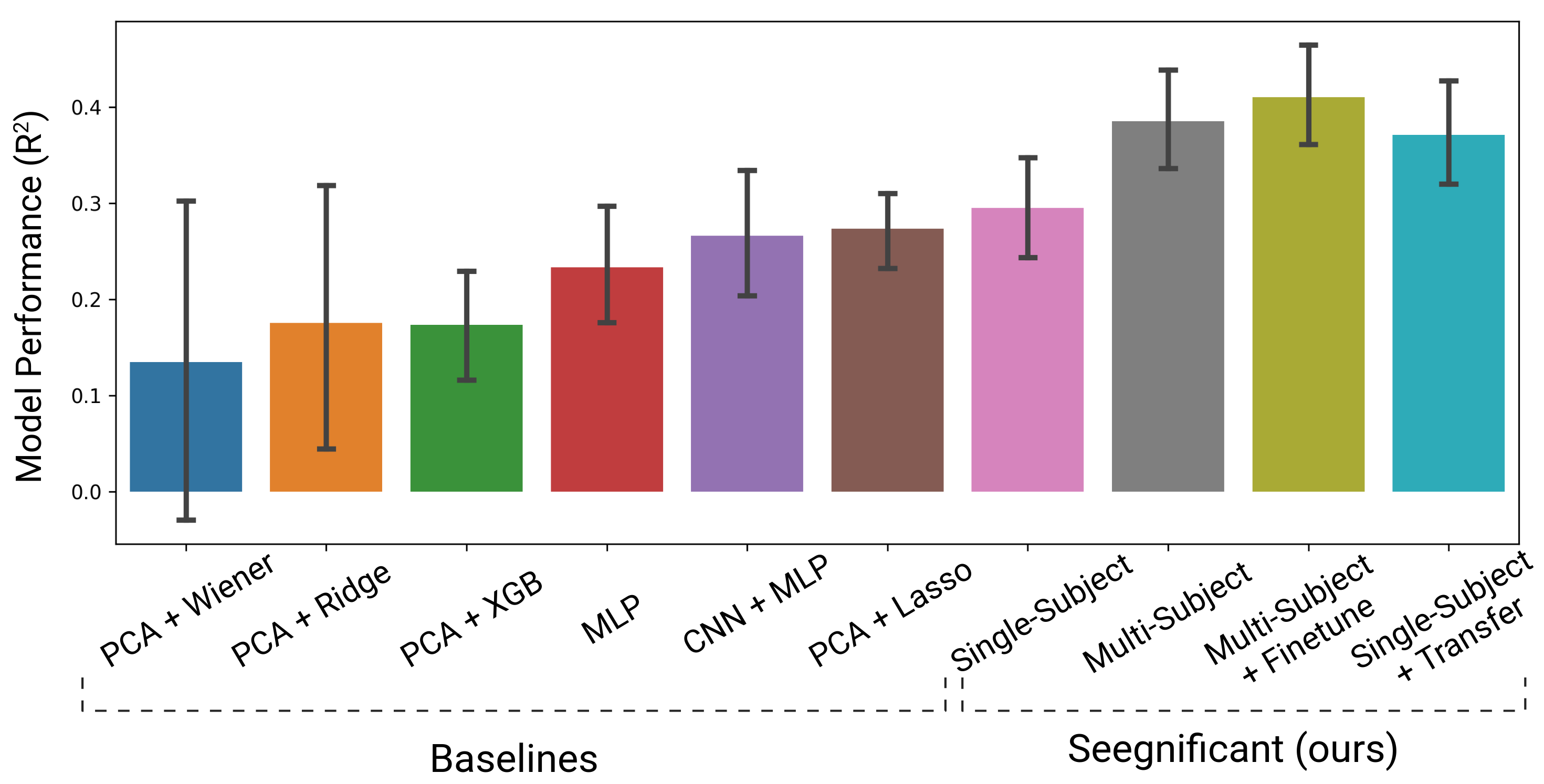} 
    \caption{\textit{{Decoding performance (mean $\pm$ sem) for various baselines and our proposed models.}}}
    \label{fig:baseline_comparisons}
\end{figure}

We were also interested in quantifying the performance benefit of our modeling approach compared to other traditional and state-of-the-art neural decoding approaches. To do so, we trained single-subject baseline models (described in detail in \ref{sec:baseline_details}), and compared those with seegnificant. We observed that seegnificant outperformed the baseline models when trained on single subject, multiple subjects, and when transferred (see Fig. \ref{fig:baseline_comparisons}). Specifically, our single-subject models (section \ref{sec:ss_training}) outperformed all baselines by an average per-subject test set $\Delta R^{2} \geq 0.03$. Our multi-subject models (section \ref{ms_training}) outperformed the baselines by a per-subject test set $\Delta R^{2} \geq 0.12$, which could be further increased to $\Delta R^{2} \geq 0.14$ by finetuning to single subjects. Importantly, our transfer-learned single-subject models (section \ref{transfering_to_new_subjects}) also outperformed all baseline models by a per-subject test set $\Delta R^{2} \geq 0.11$, on average.

\subsection{{Ablations}} 

Last, we sought to identify the contribution of each building block of our proposed architecture to the decoding performance of our models. To investigate this, we performed an ablation study on our proposed architecture when trained on multiple subject (same as in section \ref{ms_training}). To ensure a fair comparison, all training hyperparameters were kept identical across all training runs.

\subsubsection{{Model components}} \label{section:ablations}

\begin{wrapfigure}{r}{0.65\textwidth}
    \centering
    \includegraphics[scale=0.37]{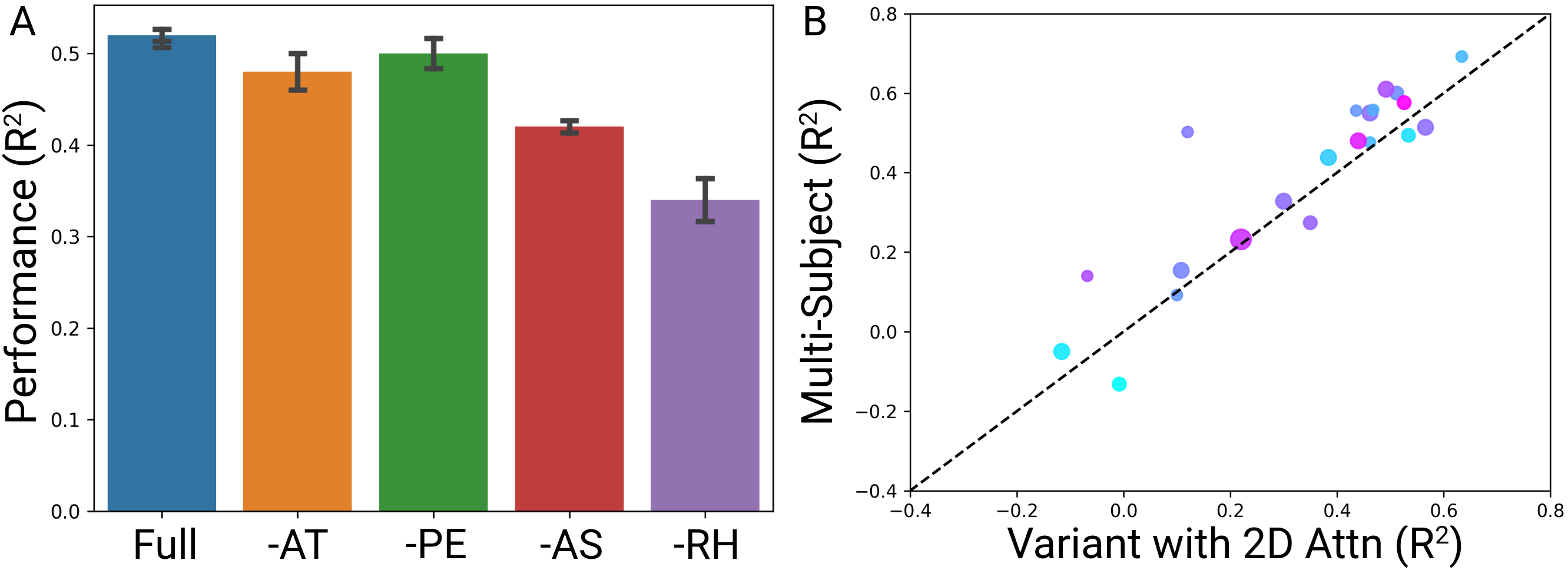} 
        \caption{{\textit{Summary of ablation results.} (A) Decoding performance (mean $\pm$ sem) across different model variants. (B) Head-to-head comparison of the decoding performance of our multi-session, multi-subject model vs its 2D attention variant.}}
    \label{fig:ablations}
\end{wrapfigure}

In this subsection, we trained our proposed model while ablating its components one by one. The model is composed of: 1. convolutional tokenizer (CT), 2. attention in time dimension (AT), 3. spatial positional encoding (PE), 4. attention in space dimension (AS), 5. multi-layer perceptron (MLP), 6. subject-specific regression heads (RH) (see also Fig. \ref{networkArchitectrure}). 
Since there is no straightforward way to process the raw sEEG data with AT without first extracting features, we did not ablate CT. Additionally, since there is no straightforward way to project the output of the AS to the regression heads without a linear layer, we also did not ablate the MLP.

The ablation study results are summarized in Fig. \ref{fig:ablations}A. Inspection of the results suggests that the subject-specific regression heads are key to the model's decoding performance ($\Delta R^{2} = 18$), likely because the response time profiles across subjects are quite diverse. Ablating the attention mechanism in the space dimension also substantially degrades decoding performance ($\Delta R^{2} = 0.10$) suggesting that it might be the main mechanism by which the model is capable of dealing with the heterogenous number/placement of electrodes across subjects.

\subsubsection{{Model variants}} \label{section:2D_attn_variant}

\begin{wraptable}{r}{0.5\textwidth}
\caption{\textit{Decoding performance (mean $\pm$ sem) for the different models}. }
\label{tab:2D_attn}
\begin{tabular}{ccc} \\\toprule  
  Model & Per-Subject $R^2$ & Train time (mins) \\\midrule   
 Variant & 0.33 $\pm$ 0.05 & 141.6 $\pm$ 3.23 \\\midrule
 Ours & 0.39 $\pm$ 0.05 & 25.7 $\pm$ 0.03 \\\bottomrule
\end{tabular}
\end{wraptable}

We also compared our architecture with a close variant. 
Our architecture is designed with two separate attention mechanisms over the time and space dimensions of our data (see section \ref{section:attn}), mainly to reduce computational burden
since the computational complexity of attention scales quadratically with sequence length.
To ensure that this choice did not negatively affect the model's decoding performance, we compared our architecture against a variant where the two separate attention mechanisms are replaced by a single 2D attention mechanism over the combined time and space 
dimensions of the data (to enable this, the positional encoding was ablated). 
\textit{We identified that our proposed architecture outperformed its 2D attention variant by a per-subject, test set $\Delta R^{2} = 0.06$, on average} (see Table \ref{tab:2D_attn} and Fig. \ref{fig:ablations}B). We also identified that our proposed architecture trained $\sim 5.5 \times $ faster.

\section{Discussion} \label{sec:discussion}
In this work, we introduce seegnificant: a novel framework and architecture that can be used for multi-session, across subject decoding based on sEEG. We show that training on diverse and large cohorts of sEEG data, despite inter-subject differences in terms of electrode number/placement, is not only possible but also leads to better decoding performance compared to single-subject approaches. We also show that the neural representations learned by pretrained models can be efficiently transferred to new subjects, despite the small number of available trials within-subjects (which is the case in any real-world clinical setting). In fact, this few-shot, transfer learning approach gives a clear benefit to training from scratch and can reach decoding performance levels that are very close to those of multi-subject pretrained models, with a significant computational benefit. 

Our results suggest that training on diverse multi-subject sEEG datasets boosts decoding performance compared to single-subject approaches. While our dataset is large compared to other works \citep{Angrick2021, Petrosyan2022, Meng2021, Wu2022, Wu2024}, scaling up further will inevitably require combining data collected while subjects perform a variety of different behavioral tasks. Advances in other fields have shown that training on multi-task data can improve performance and generalization \citep{Azabou2023, ruder2017}. Therefore, investigating whether those performance/generalization gains will hold on sEEG decoding is worth exploring. Training on multi-task datasets would also give insight as to which aspects of neural computations are shared across tasks and which are not. We leave the investigation of multi-task model training as future work.

Given the clinical circumstances under which sEEG is recorded, performance to downstream decoding tasks would likely benefit from pretraining using self-supervised objectives. When patients are admitted in the epilepsy monitoring unit, their sEEG recordings are collected continuously over 5-10 days. In contrast, sEEG study participants perform only a few minutes of a behavioral task. Evidently, the amount of unsupervised data collected is orders of magnitude more than the supervised data. Strategies such as autoencoding, masked modeling, or next token prediction could enable training on the vasts amounts of data that are collected but unused by supervised learning approaches. 

Overall, this work advances clinical and human neuroscience research by introducing a framework that enables multi-subject model training based on sEEG. We show that models pretrained on large and diverse sEEG datasets can be easily transferred to new subjects and therefore can be used to accelerate progress in the neural decoding domain. Given that sEEG is currently the gold standard invasive neural recording modality used in humans, our work advances progress towards sEEG-based neural decoding therapeutic interventions, ultimately bringing them closer to clinical translation.

\acksection

We would like to thank the following funding sources for their generous support: 
National Institutes of  Health grant R01NS121219 (FV);
Office of Naval Research grant N00014-22-1-2677 (KD);
National Institutes of Health grant 6T32NS091006 (AGR);
Onassis Foundation graduate student scholarship and A. G. Leventis Foundation graduate student scholarship (GM). Authors declare no competing interests.

\addcontentsline{toc}{chapter}{BIBLIOGRAPHY}
\bibliographystyle{abbrvnat}
\bibliography{bibliography}

\newpage
\appendix

\section{Appendix / supplemental material}

\subsection{Data collection and signal processing details} \label{data_collection_details}

\textbf{Experimental setup}. Stereotactic EEG was recorded from 23 participants diagnosed with medically refractory epilepsy who underwent surgical implantation of intracranial electrodes for seizure localization. Patients were implanted with intraparencymal depth electrodes (“stereo EEG,” Ad-tech, 1.1 diameter, 4 contacts spaced 5 mm apart), except in one patient who also had subdural grid electrodes (Ad-tech, 4 mm contacts, spaced 10 mm apart). sEEG was recorded using a Natus recording system. Based on the amplifier and the discretion of the clinical team, signals were sampled at either 512 or 1024 Hz. Clinical circumstances alone determined the number and placement of the implanted electrodes. To participate in this study, participants provided written informed consent in accordance with the IRB of the University of Pennsylvania and were compensated for their time. 

\textbf{Behavioral task}. Participants performed the color-change-detection-task, a stimulus-detection task with a variable foreperiod delay \citep{Buch2022, Ramayya2022} while their sEEG activity was simultaneously recorded. Participants viewed visual stimuli on a laptop and responded by pressing a button on a game controller. The behavioral task is composed of sequences of trials described here. Each trial began with the presentation of a visual stimulus (small white box) at a randomized location on the screen as a fixation target (one of nine locations on a 3 x 3 grid). The stimulus changed color (from white to yellow) after one of two randomly selected foreperiod delays: (i) 500 ms, or (ii) 1500 ms. The response time (time between color change and button press) for each trial was recorded. 

\textbf{Signal preprocessing}.
The sEEG signals were first converted to a bipolar montage by taking the difference between pairs of immediately adjacent contacts on the same electrode shank. The resulting  bipolar signals were treated as new virtual electrodes (henceforth, electrodes) whose virtual location was the midpoint between the adjacent contacts \citep{Ramayya2022, Burke2013}. Within electrodes, we computed the line noise SNR, as the ratio of each electrode's spectral power in the frequency range of 58-62 Hz over the spectral power in the range 18-22 Hz.  Electrodes for which the SNR was greater than one were excluded from further analysis. Additionally, disconnected electrodes, defined as electrodes whose voltage had a standard deviation of zero were also excluded from further study. For the remaining electrodes, we investigated for noise contaminated trials defined as trials where (i) the voltage was not recorded due to saturation of the amplifier, or (ii) the mean $\pm$ SD of the voltage of the trial was greater than 10 times the mean of all trials of that electrode. Trials that were determined as noise contaminated were then removed.

\textbf{Signal processing details}. To combat the heterogeneity of electrode placement between subjects, we identified the subset of electrodes across subjects that responded to the stimulus color change while participants performed the color-change-detection-task. To do so, we followed the methodology used by \cite{Paraskevopoulou2021}. Specifically, we identified electrode locations where high-$\gamma$ band activity was significantly modulated by the stimulus color change.

We first filtered the data, separately for each electrode, to the high-$\gamma$ band (70-150 Hz) using a 4\textsuperscript{th} order Butterworth filter and subsequently extracted the high-$\gamma$ envelope using the Hilbert Transform. Then, we epoched the high-$\gamma$ envelopes around the color-change ([-500, 1500] ms) and separated each trial into a baseline period ([-500, 0] ms, before the color-change) and a task period ([0, 1500] ms, after the color-change).
For each electrode, we computed the median high-$\gamma$ amplitude time course of the task ($Median_{task}$) and baseline ($Median_{baseline}$) periods across trials. Then for each electrode, we calculated the signal-to-noise ratio (SNR) as 
\begin{equation}
    SNR = \frac{\sigma^{2}(Median_{task})}{\sigma^{2}(Median_{baseline})},
\end{equation}
where $\sigma^{2}(x)$ refers to the variance of $x$. To identify electrodes with statistically significant SNR values, we applied a bootstrap randomization test in which we randomly shuffled the task and baseline labels from all locations 10,000 times and computed one random SNR value for each such iteration. We then calculated p-values for each electrode location as the fraction of randomized SNR values that were larger than the computed SNR. We determined responsive electrodes as those with a p-value $< 0.05$, which corresponds to a confidence level $\alpha = 0.05$. Prior to determining significance, we adjusted the p-values calculated across all electrodes of all subjects using false discovery rate \citep{Benjamini1995}. 

The \textit{broadband} voltage traces of all responsive electrodes were then z-scored (within electrodes), and downsampled to 400 Hz. Those electrodes were then used for decoding response times by our models. We note that using this procedure, we also rejected two participants from the study for whom only zero and one electrodes were returned as significant (lowering our participant count from 23 to 21). This electrode selection procedure lowered our computational cost, and because it merely uses high-$\gamma$ to select electrodes, it avoids problems of statistical double dipping.

\subsection{Model training details} \label{ModelTrainingDetails}

\textbf{Model complexity.} The total number of trainable parameters for the multi-session, multi-subject model is 797,095. From those, 753,394 are shared across subjects (shared trunk) and the rest 43,701 parameters are subject-specific (parameters of the regression heads with 2,081 parameters per subject).

\textbf{Training hyperparameters}. All models were implemented and trained using Pytorch 2.1.0+cu121 \citep{Pytorch}. AdamW was used as the optimizer \citep{Leszczyski2020} (with $b_{1} = 0.5$ and $b_{2} = 0.999$). All models were trained for 1000 epochs. A step learning rate scheduler was used with an initial learning rate set to $10^{-3}$ and decayed by a factor of 0.5 every 200 epochs for single-subject models and by a factor of 0.9 every 100 epochs for multi-subject models. Batch size was fixed to 64 and 1024 for all single-subject and multi-subject models, respectively. {All models were optimized using Huber loss, except for when finetuning the multi-session, multi-subject model (see section \ref{ms_training}) to individual subjects, where MSE loss was used.} 

{\textbf{Hyperparameters of the spatial positional encoding.} For all models, the gaussian kernels used in the spatial positional encoding were centered at $ \mu \in \{-90, -70, -50, \dots, 50, 70 \}$ and had variances $\sigma^{2} \in \{1, 2, 4, \dots, 64 \}$.}

\textbf{Controlling for random data splits}. {For all models, the train/validation/test split was 70/15/15 \%.} To ensure that our results would not be biased in any way by the data splits, experiments performed in sections \ref{sec:ss_training}, \ref{ms_training},{ \ref{sec:baseline_comparisons}, \ref{section:2D_attn_variant}, \ref{section:positional_encoding}, and \ref{section:penc_comparisons}} were run using 5 different data splits and the results were averaged across the different splits. For experiments described in section \ref{transfering_to_new_subjects}, multi-subject models were trained for 1 data split (since training 21 multi-session, multi-subject models $\times$ 5 times for the different data split would be very computationally expensive). Finetuning to single-subjects was performed across 5 different data splits and the results across the splits were averaged. For the ablation study described in section \ref{section:ablations}, models were trained for 3 data splits, to reduce computational burden.

\textbf{Finetuning.} For finetuning multi-session, multi-subject models to single subjects in section \ref{ms_training}, all model weights were updated  throughout the single-subject finetuning. For transferring multi-session, multi-subject models to single subjects in section \ref{transfering_to_new_subjects}, since the regression heads of the left-out participant were completely untrained, we gradually unfroze model weights. For the first 400 training epochs, only the regression head {(2,081 parameters per subject)} of the left-out participant was trained {(the remaining 753,394 model parameters were frozen)}. For the remaining 600 epochs, all model parameters were trained. All training parameters were identical to those described in paragraph "Training hyperparameters" above.

\textbf{Computational resources.} All models were trained on a machine with an AMD EPYC 7502P 32-Core Processor and 1 Nvidia A40 GPU with 44.99 GiB of memory. Single-subject models trained on average within 5 mins and multi-subject models trained within an hour. {The memory requirement to train the multi-session, multi-subject model with a batch size of 1024 was $\sim$ 8 GiB.}

\subsection{Baseline model details} \label{sec:baseline_details}

{\textbf{Non deep learning models.} The following non-deep learning models were trained on single-subjects, using scikit-learn (version 1.2.2): 1. Wiener Filter, 2. Ridge Regression, 3. Lasso Regression, 4. Gradient-Boosting (XGB). All models were used as part of a pipeline composed of: standard scaling $\rightarrow$ PCA $\rightarrow$ model. For Ridge and Lasso Regression, we optimized over the hyperparameter \texttt{alpha} $\in \{0.1, 1, 10\}$. For XGB, we optimized over the hyperparameter \texttt{n\_estimators} $\in \{50, 100, 200\}$. All other hyperparameters were kept at their default values. }

{\textbf{Deep learning models.}
The following deep learning models were trained on single-subjects, using the training procedure described in section \ref{ModelTrainingDetails} (same as that used for our proposed architecture): 1. MLP (4 layers), 2. CNN + MLP (1 + 3 layers, respectively).}

\subsection{Additional experiments}

\subsubsection{Model real-time applicability} \label{section:RT_applicability}

{To understand whether our multi-session, multi-subject model (see section \ref{ms_training}) could be used in real time, which is essential for any real-world neural decoding system, we measured our model's inference time on two machines: a commercial laptop and a server. \textit{The results, summarized in Table \ref{tab:inference_times}, show that our model runs in $< 10$ msec on both machines, on either CPU or GPU}. This indicates that our proposed architecture can be easily used in real time, since real-time systems run on the order of 100 msec.}

\begin{table}[h!]
\begin{center}
\caption{\textit{Model inference time on different hardware. Units are in msec.}}
\label{tab:inference_times}
\begin{tabular}{ccc} \\\toprule  
  \textbf{Machine} & \textbf{CPU} & \textbf{GPU} \\\midrule  
  AMD EPYC 7502P + Nvidia A40 & 9.1 & 5.1 \\\midrule   
 Intel Core i9 + Nvidia A2000 & 4.0 & 7.9 \\\bottomrule
\end{tabular}
\end{center}
\end{table}

\subsubsection{Contribution of spatial positional encoding} \label{section:positional_encoding}

Following our ablation study (see section \ref{section:ablations}), we sought to better understand the contribution of our spatial positional encoding to the model's performance. Contrary to other positional encoding techniques, constructed to encode the order of tokens in a sequence (see \cite{Vaswani2017} for Fourier and \cite{Su2024} for rotation based approaches), our approach encodes tokens with their 3D location in the brain, based on each electrode's MNI coordinates.

\begin{wrapfigure}{r}{0.4\textwidth}
    \centering
    \includegraphics[scale=0.45]{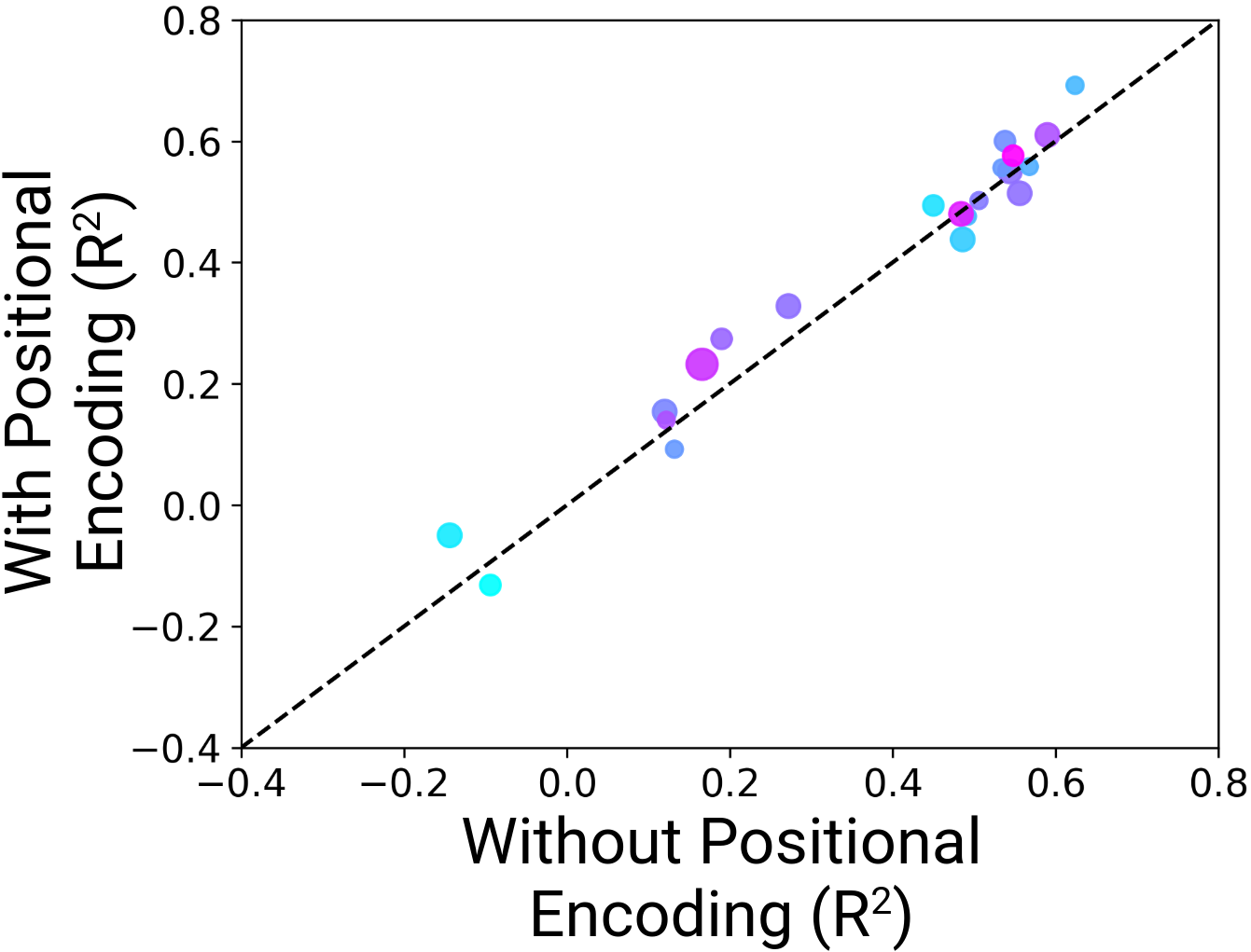} 
    \caption{\textit{Per subject decoding performance comparison between the model with and without spatial positional encoding}.}
    \label{fig:penc_contributions}
\end{wrapfigure}

To quantify the performance gains, we trained a variant of our multi-session, multi-subject model, which was identical to that described in section \ref{ms_training} with the positional encoding ablated. 

We observed that \textit{ablating the the positional encoding leads to a per-subject decoding performance decrease of $\Delta R^{2} = -0.02$, on average}. We show a within-subject head to head comparison of the performance of the two models in Fig. \ref{fig:penc_contributions}.

We also investigated whether the decoding performance of the model with and without positional encoding is significantly different. Specifically, we performed a Wilcoxon rank-sum test between the groups: (1) test set $R^{2}$ scores of all subjects obtained by training the multi-subject model with spatial positional encoding, and (2) test set $R^{2}$ scores of all subjects obtained by training the multi-subject model without spatial positional encoding. The test returned a test-statistic = 0.34 and a p-value = 0.73, indicating that there is no significant difference between the performance of the model with and without spatial positional
encoding.

{Those results, while not significant, suggests that informing electrode-tokens with their 3D location in the brain prior to self-attention in the space dimension might slightly benefit decoding performance. Intuitively, this would makes sense, since the functional computations happening within roughly the same brain regions are similar across humans. }

\subsubsection{Comparison of our proposed spatial positional encoding against other approaches.} \label{section:penc_comparisons}

{The results of section \ref{section:positional_encoding} suggest that the decoding performance gains introduced by the positional encoding scheme used in this work are only modest. Therefore, we explored whether other positional encoding schemes would boost decoding performance further. To do so, we trained variants of our proposed architecture on the combined data of all subjects (same training scheme as the one used in section \ref{ms_training}) with the positional encoding replaced by the: 1. positional encoding scheme introduced by \cite{Vaswani2017}, 2. a variant of the positional encoding scheme used in this work where RBFs are replaced by fourier functions. While the explored schemes are far from comprehensive, we identified that our proposed positional encoding scheme performed as well as or better than other approaches (see Table \ref{tab:penc_comparisons}). However, the decoding performance gains provided by all positional encodings explored in this work are modest. Therefore, it would be worth exploring other positional encoding schemes, perhaps based on whole brain MRI images and/or brain atlases other than MNI \citep{talairach1988}.}

\begin{table}[h!]
\begin{center}
\caption{\textit{Model performance using different positional encoding schemes.}}
\label{tab:penc_comparisons}
\begin{tabular}{lc} \\\toprule
  Positinal Encoding & Per-Subject $R^{2}$ \\\midrule
 \cite{Vaswani2017} & 0.16 $\pm$ 0.04 \\\midrule
 MNI-Fourier & 0.39 $\pm$ 0.05 \\\midrule
 MNI-RBF (Ours) & 0.39 $\pm$ 0.05\\\bottomrule
\end{tabular}
\end{center}
\end{table}

\newpage
\section*{NeurIPS Paper Checklist}

\begin{enumerate}

\item {\bf Claims}
    \item[] Question: Do the main claims made in the abstract and introduction accurately reflect the paper's contributions and scope?
    \item[] Answer: \answerYes{} 
    \item[] Justification: To the best of our knowledge, all claims made in the abstract and introduction are supported by our experiments described in section \ref{sec:experiments}.

\item {\bf Limitations}
    \item[] Question: Does the paper discuss the limitations of the work performed by the authors?
    \item[] Answer: \answerYes{} 
    \item[] Justification: We discuss the limitations of our proposed approach in sections \ref{sec:approach} and \ref{sec:discussion}. 

\item {\bf Theory Assumptions and Proofs}
    \item[] Question: For each theoretical result, does the paper provide the full set of assumptions and a complete (and correct) proof?
    \item[] Answer: \answerNA{} 
    \item[] Justification: No theoretical results are introduced in this work.

    \item {\bf Experimental Result Reproducibility}
    \item[] Question: Does the paper fully disclose all the information needed to reproduce the main experimental results of the paper to the extent that it affects the main claims and/or conclusions of the paper (regardless of whether the code and data are provided or not)?
    \item[] Answer: \answerYes{} 
    \item[] Justification: We describe the methodology introduced in this work in section \ref{sec:approach}. The details to reproduce the behavioral experiment used in this work are provided in section \ref{data_collection_details}. The details of how to train the models introduced in this work are provided section \ref{ModelTrainingDetails}. 

\item {\bf Open access to data and code}
    \item[] Question: Does the paper provide open access to the data and code, with sufficient instructions to faithfully reproduce the main experimental results, as described in supplemental material?
    \item[] Answer: \answerNo{} 
    \item[] Justification: Our code is available at https://github.com/gmentz/seegnificant. The dataset used in this work contains electrophysiological data collected from human patients. To respect their privacy and comply with HIPAA regulations, we cannot make the dataset public. 

\item {\bf Experimental Setting/Details}
    \item[] Question: Does the paper specify all the training and test details (e.g., data splits, hyperparameters, how they were chosen, type of optimizer, etc.) necessary to understand the results?
    \item[] Answer: \answerYes{} 
    \item[] Justification: We refer the reader to section \ref{ModelTrainingDetails}.

\item {\bf Experiment Statistical Significance}
    \item[] Question: Does the paper report error bars suitably and correctly defined or other appropriate information about the statistical significance of the experiments?
    \item[] Answer: \answerYes{} 
    \item[] Justification: All figures with barplots include error bars and the factors of variability capture by them are included in captions of the figure. One statistical test (corrected appropriately for multiple comparisons) was used in this work, described in section \ref{data_collection_details}.

\item {\bf Experiments Compute Resources}
    \item[] Question: For each experiment, does the paper provide sufficient information on the computer resources (type of compute workers, memory, time of execution) needed to reproduce the experiments?
    \item[] Answer: \answerYes{} 
    \item[] Justification: We refer the reader to \ref{ModelTrainingDetails}.
    
\item {\bf Code Of Ethics}
    \item[] Question: Does the research conducted in the paper conform, in every respect, with the NeurIPS Code of Ethics \url{https://neurips.cc/public/EthicsGuidelines}?
    \item[] Answer: \answerYes{} 
    \item[] Justification: To the best of our knowledge, this manuscript conforms in every aspect with the NeurIPS Code of Ethics.

\item {\bf Broader Impacts}
    \item[] Question: Does the paper discuss both potential positive societal impacts and negative societal impacts of the work performed?
    \item[] Answer: \answerYes{} 
    \item[] Justification: Please refer to \ref{sec:introduction}, and \ref{sec:discussion}. 
    
\item {\bf Safeguards}
    \item[] Question: Does the paper describe safeguards that have been put in place for responsible release of data or models that have a high risk for misuse (e.g., pretrained language models, image generators, or scraped datasets)?
    \item[] Answer: \answerNA{} 
    \item[] Justification: The data used in this work are not shared to protect the privacy of the human participants whose electrophysiological data was used in this study. To the best of our knowledge, the code shared with this work does not pose a high risk for misuse.

\item {\bf Licenses for existing assets}
    \item[] Question: Are the creators or original owners of assets (e.g., code, data, models), used in the paper, properly credited and are the license and terms of use explicitly mentioned and properly respected?
    \item[] Answer: \answerYes{} 
    \item[] Justification: We provide proper citations for all the works upon which this work is built. 

\item {\bf New Assets}
    \item[] Question: Are new assets introduced in the paper well documented and is the documentation provided alongside the assets?
    \item[] Answer: \answerYes{} 
    \item[] Justification: All details of the code are discused in section \ref{ModelTrainingDetails}. Code is publicly available at https://github.com/gmentz/seegnificant.

\item {\bf Crowdsourcing and Research with Human Subjects}
    \item[] Question: For crowdsourcing experiments and research with human subjects, does the paper include the full text of instructions given to participants and screenshots, if applicable, as well as details about compensation (if any)? 
    \item[] Answer: \answerNo{} 
    \item[] Justification: We do not have access to the full text of instructions given to participants. 

\item {\bf Institutional Review Board (IRB) Approvals or Equivalent for Research with Human Subjects}
    \item[] Question: Does the paper describe potential risks incurred by study participants, whether such risks were disclosed to the subjects, and whether Institutional Review Board (IRB) approvals (or an equivalent approval/review based on the requirements of your country or institution) were obtained?
    \item[] Answer: \answerYes{} 
    \item[] Justification: Approval was obtained by the IRB of the University of Pennsylvania prior to data collection.

\end{enumerate}

\end{document}